\documentclass{article}

\usepackage{graphicx}
\usepackage{arxiv}

\usepackage[utf8]{inputenc} 
\usepackage[T1]{fontenc}    
\usepackage{hyperref}       
\usepackage{url}            
\usepackage{booktabs}       
\usepackage{amsfonts}       
\usepackage{nicefrac}       
\usepackage{microtype}      
\usepackage{lipsum}

\author{Letizia Jaccheri\\
Norwegian University of Science and Technology (NTNU)\\
Trondheim\\
Norway\\
\texttt{Letizia.Jaccheri@ntnu.no}\\

   \And
   Sandro Morasca\\
Universit\`{a} degli Studi dell'Insubria\\
Varese\\
 Italy\\
  \texttt{Sandro.Morasca@uninsubria.it}\\
  
  }

\title{Toward Inclusion of Children as Software Engineering Stakeholders}

\begin{document}

\maketitle

\begin{abstract}
Background: A growing amount of software is available to children today. Children use both software that has been explicitly
developed for them and software for general users. While they obtain clear benefits from software, such as access to creativity tools
and learning resources, children are also exposed to several risks
and disadvantages, such as privacy violation, inactivity, or safety
risks that can even lead to death. The research and development
community is addressing and investigating positive and negative
impacts of software for children one by one, but no comprehensive model exists that relates software engineering and children as
stakeholders in their own right.
Aims: The final objective of this line of research is to propose
effective ways in which children can be involved in Software Engineering activities as stakeholders. Specifically, in this paper, we
investigate the quality aspects that are of interest for children, as
quality is a crucial aspect in the development of any kind of software, especially for stakeholders like children.
Method: Our contribution is based mainly on an analysis of
studies at the intersection between Software Engineering (especially
software quality) and Child Computer Interaction.
Results: We identify a set of qualities and a preliminary set of
guidelines that can be used by researchers and practitioners in
understanding the complex interrelations between Software Engineering and children. Based on the qualities and the guidelines,
researchers can design empirical investigations to obtain deeper insights into the phenomenon and propose new Software Engineering
knowledge specific for this type of stakeholders.
Conclusions: This conceptualization is a first step towards a
framework to support children as stakeholders in software engineering.

\end{abstract}

\section{Introduction}


A massive and ever growing amount of software intensive technologies  is available today to children of younger and younger age through sites such as Facebook and Instagram, apps,  games, and Internet of Things (IoT) devices embedded in, for example, cars and toys.  In some cases, the software is specifically made for children, but, in other cases, it is made for the general users, like Facebook and Instagram, and also used by children. In some cases, the software aims at solving a problem, like helping children with reading difficulties to learn how to read \cite{holmes2011using} or children with obesity to exercise \cite{hagen2016gameplay}. In other cases, the software's goal is to enhance children creativity \cite{papavlasopoulou2017empirical}, like Scratch \cite{resnick2009scratch}. In most cases, software is made for commercial and entertainment purposes, like war games.

With the advancement of technology and the development of new IoT  applications, games, and social media sites, it is getting increasingly difficult to keep up with the associated threats and vulnerabilities for all stakeholders, and especially for children. 
The increasing presence of robotics, automated systems, and AI makes software more and more pervasive at all levels and for all ages. This has ethical and social implications for software engineers and users, especially for a particularly vulnerable category of users like children \cite{unicef2019}.
In the last few years, we have witnessed a series of problems generated in the digital ecosystem populated by software and children.  
Children can fall victim to cybersecurity threats like social engineering, cyber bullying, hacking, viruses, and damaging malware, cyber stalking, etc. through search engines, online advertisements and social networking websites such as Facebook, Twitter and lots of other websites \cite{hamdan2013protecting}.
The US National Safety Council, which tracks hot car deaths across the United States, reports than an average of 38 children under 15 die each year of heatstroke related to being trapped in a hot car --– whether due to caregiver error or a child climbing into a vehicle and being unable to escape. Children's safety apps like ``Kars 4 Kids'' have been developed and are available to parents who want to eliminate these risks. 
Solutions that have been not specifically been designed for children, like airbags, pose serious risks for children.
Available data indicate that, on average, children under
the age of 13 are more likely to be harmed by an
airbag than to be helped by it \cite{worldreport}. Mental health issues and even suicide are been caused by interaction with software. Instagram removed nearly 10,000 images related to suicide and self-harm every day in the months following the Molly Russell scandal (according to the Telegraph newspaper). 
Physical health issues, like obesity \cite{vandewater2004linking} and diabetes have been linked to the excessive use of videogames. 

Given the greater exposure of children to digital technologies, the interaction design research community take into account children's abilities, interests, and developmental needs \cite{hourcade2008interaction} when designing interfaces of software for children. 
International organizations that focus on
Child’s Rights, such as UNICEF \cite{unicef2019} have recently started to examine the emerging ethical considerations regarding software development for children. 

We observe that the community is producing new software for children to address the problems generated by software without considering the problems that the new software will produce. For example, \cite{hagen2016gameplay} reports about  development and evaluation of an exergame based on a shooting game, to improve physical well-being without considering how the other qualities (for example mental well-being, or privacy) will be impacted by this new game.   
This is a classical trade-off in software engineering: when optimizing one quality attribute, for example performance, one needs to be careful not to compromise other qualities, like for example readability and maintainability.

The final research objective of the line of research of our paper is the development of new knowledge which will enable to recognize children as competent stakeholders in Software Engineering.

Existing quality models, e.g., \cite{ISO25010:2011} already deal with some of the qualities that are important for children, e.g., security. However, it is not straightforward to figure out whether and to what extent they also deal with qualities like creativity or well-being. 
It is therefore necessary to build a comprehensive quality model for software development when children are stakeholders. This quality model should be used for several purposes: (1) assess the overall quality of the final software product; (2) assess the quality of the various artifacts produced during software development; (3) allow software stakeholders to make informed decisions about the inevitable trade-offs between all of the relevant qualities during the development of a software product for children; (4) allocate resources in an effective way to reach quality objectives with the available budget.

In this paper, we introduce an initial set of qualities that are relevant to children as Software Engineering stakeholders. 
The main qualities that we propose are: Security, Well-being, Fun, and Creativity. 
We also provide a preliminary set of guidelines for researchers and practitioners that can help them  understand the complex interrelations between Software Engineering and children. Starting from the qualities and the guidelines, researchers can design empirical investigations to obtain deeper insights into the phenomenon and propose new software engineering knowledge specific for this type of stakeholders. Practitioners can use the provided knowledge to better understand children as stakeholders of their software products. 



The remainder of this paper is organized as follows. 
Section \ref{s:SoftwareQualityModels} presents the relevant background on software quality and quality models. 
In Section \ref{softwareengineeringchildren}, we propose a set of qualities that form a preliminary software quality model when children are primary stakeholders. In Section \ref{guidelines}, we propose preliminary guidelines for research and practice. 
Section \ref{conclusions} provides conclusions and an outline for further work.

\section{Software Quality Models} 
\label{s:SoftwareQualityModels}

A number of different process and product qualities have been identified as relevant for SE practice and research and a large number of measures have been proposed in the literature for their assessment. 

Qualities are traditionally divided into \textit{internal} qualities, such as software size, structural complexity, cohesion, and coupling, and \textit{external} qualities, such as reliability, usability, and performance \cite{FentonBiemanBOOK2014}. Internal software qualities refer to a software product or process \textit{per se}. External software qualities refer to a software product or process and to its users/stakeholders. For instance, the size of some software product depends only on the product itself, while its reliability depends on the product itself \textit{and} the way the product is used.

The distinction between internal and external qualities has practical consequences. Internal qualities have no practical interest \textit{per se}, while external qualities are the relevant ones from a practical point of view. The number of lines of code of a software product is simply a statistic, while an assessment of its reliability (e.g., how often it fails) is useful to developers and users/stakeholders.

However, this is not to say that internal measures are useless. An internal measure has practical value if a model exists that relates it to an external quality, i.e., if a model exists that can be used to quantify/estimate/predict an external quality of practical interest \cite{MorascaESEM2009}. For instance, the number of lines of code is an useful size measure because it is used in several models for various external qualities, such as reliability, fault-proneness, and reusability.

Software quality models (e.g., those of the SQUARE 25000 series \cite{ISO25010:2011}) provide an organized view of a number of qualities that are believed to be important in the evaluation of software products (and processes). Quality models are usually general-purpose, in that their objective is to take into account the needs and goals of many and diverse software users and developers. However, to be practically used, quality models need to be ``instantiated'' for specific sectors, or companies, or even projects.

The fact that it was promoted from being a subcharacteristic of functionality in the ISO/IEC 9126-1:2001 standard to being a full-fledged characteristic in the ISO/IEC 25010:2011 standard shows that Security has become a fundamental quality for all types of software over the years. Security has five subcharacteristics in the ISO/IEC 25010:2011 standard: 1) confidentiality, i.e., allowing only authorized actors to access data; 2) integrity of software or its data; 3) non-repudiation, i.e., proof of the occurrence of actions or events that have taken place; 4) accountability, e.g., traceability of actions, such as transactions; 5) authenticity, i.e., identifiability of the actors interacting with software.

The ISO/IEC 25010:2011 standard also includes a ``Quality in Use'' model, which includes five characteristics, which are about how a software product interacts with its stakeholders, namely, 1) Effectiveness, 2) Efficiency, 3) Satisfaction, 4) Freedom from Risk, and 5) Context Coverage. Satisfaction is refined into Usefulness, Trust, Pleasure, and Comfort. Freedom from Risk encompasses: Economic Risk Mitigation, Health and Safety Risk Mitigation, and Environmental Risk Mitigation. Subcharacteristic Health and Safety Risk Mitigation is the one that is most related to safety, but it is quite generic, in that it is related to potential risk to people in the context of use. We can envision physical and mental risks related to the use of software.

To the best of our knowledge, no quality models have been proposed or specifically tailored for software for children, which also balance security, the various dimensions of well-being, fun and creativity.

\section{Relevant Product Qualities for Children}
\label{s:SoftwareEngineeringAndChildren} 
\label{softwareengineeringchildren}

\label{s:Qualities}

It is therefore necessary to build a comprehensive quality model for software development when children are stakeholders
 To this end, in addition to being functionally correct, software products must have adequate levels of a number of qualities that are specifically relevant for children.

It is certainly too early to identify the quality levels (i.e., the thresholds and constraints) that need to be satisfied by software products for children. However, as a necessary preliminary step, we need to identify the qualities themselves that are of interest when children are stakeholders.

Quality models were generally introduced in such a way that they could be customized for specific application domains, for specific users, and for specific goals and needs. However, they were all probably conceived with grown-ups as stakeholders, and there is no indication that children were included in the set of stakeholders even as an afterthought. However, given the ever-increasing pervasiveness of software, children are a set of stakeholders that is becoming more and more important, since software is already affecting children's lives and will affect them more and more in the years to come.

We here propose a preliminary set of qualities that are specifically relevant for children, in several classes of applications, both those explicitly developed for children (e.g., games, etc.) and those developed for the general users (e.g., social media, etc.), but with children as stakeholders. Some qualities have already been addressed in existing quality models, but others may be missing or they may not have received sufficient emphasis.

The relation between the software and the child is bidirectional. On the one hand, it is important that the software exhibit some specified  characteristics, but on the other hand is important that the child be empowered with knowledge necessary to interact with the software so that this characteristic is achieved. For example, for security, on one hand, the software must be designed and developed so that it does not have security traps. On the other hand, each child must be empowered with knowledge and awareness about security. The same holds for fun, creativity, and well-being.

\begin{figure}
    \centering
    \includegraphics[width=1\columnwidth]{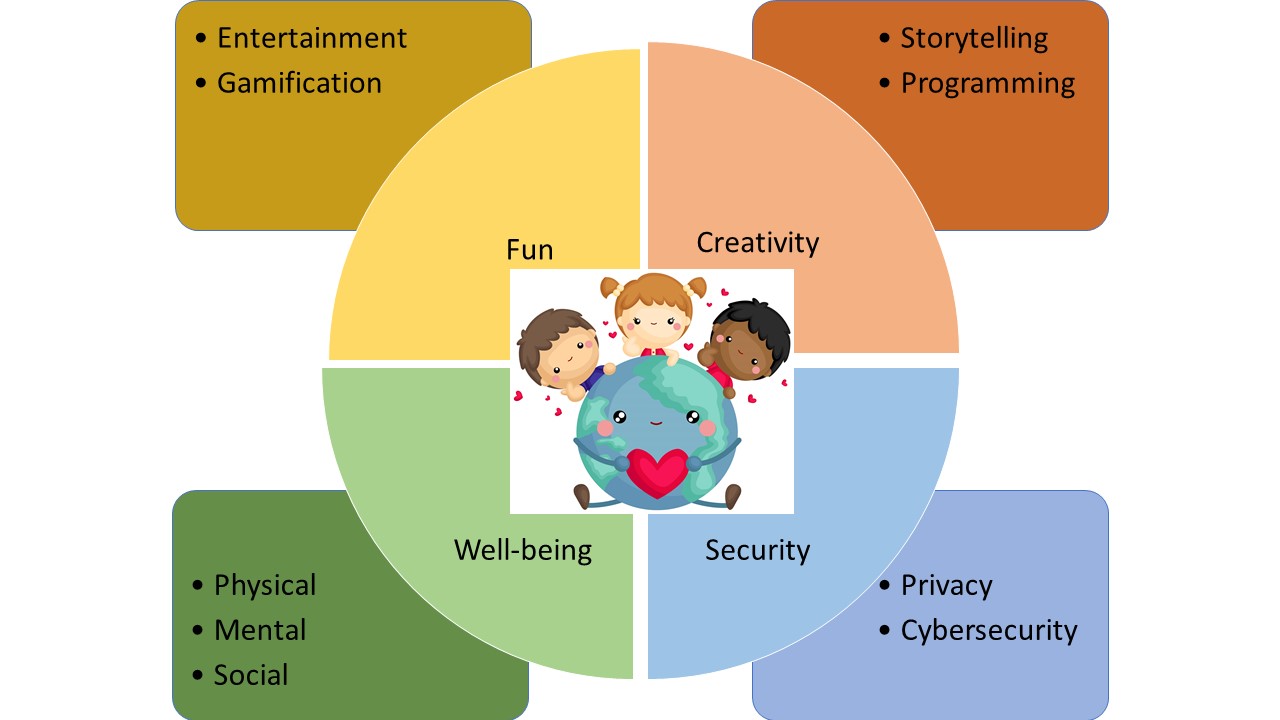}
    \caption{Preliminary Quality model for Children Software. Uses picture designed by user10320847 / Freepik.}
    \label{figure_surrounded}
\end{figure}

\subsection{Security} \label{s:Security}
Security is an important characteristics of all software products and gets even more importance when it comes to software for children.

Ensuring \textbf{secure} interaction between children and a software system entails several different sub-challenges. The two most important subcharacteristic are:

\begin{itemize}
\item \textbf{Cybersecurity} is related to all those threats that may affect teenagers and countermeasures to support teens and their parents and the awareness that teenagers have on the various cybersecurity threats. For example, \cite{hamdan2013protecting} reports an investigation about teen agers and cybersecurity awareness.  
A  mobile app called CyberAware, destined to cybersecurity education and awareness is reported in   \cite{Giannakas2016}. 
\item Privacy is related to how to ensure that private information about the children is not made public (\textbf{privacy}). Privacy is a characteristics of the software system, that has to be carefully addressed by the software engineers. How to deal with own information and how to share online, is a skill that children and teenagers have to acquire. \cite{lwin2008protecting} reports about the role of parents of influencing children's willingness to disclose information online. 
\end{itemize}
In this respect, education plays a fundamental role. It is of paramount importance to ensure that children are fully aware of the importance of protecting public data (cybersecurity) and own data (privacy). 

The ISO/IEC 25010:2011 standard's subcharacteristic that is most closely related to the above issues is Confidentiality.

\subsection{Well-being} \label{s:Well-being}
Well-being has physical, mental, and social aspects that need to be identified and addressed, as follows.
\begin{itemize}

\item \textbf{Physical} well-being is addressed for example by the studies about exergames which show how children who suffer from game addiction and obesity \cite{vandewater2004linking} may become physically active 
 \cite{hagen2016gameplay}. 
 
 In \cite{marikken-toddlers}, the authors have studied how healthcare games and applications for toddlers who suffer from respiratory issues.
 Physical well-being can also be related to safety, like for example, addressing the question of how to ensure that cyber physical systems (like robots, cars, and even digital toys) do not physically harm children safety. 
\item Concerning the \textbf{mental} dimension, \cite{michalsen2020family} reports about how to develop software for motivating adolescents with Intellectual Disabilities to become active.  

\item We define \textbf{social} well-being as the ability to establish and maintain healthy relations to other people.   \cite{gomez2019digital} introduces a digital story tool that facilitates the process of connecting human beings and increase empathy as a function of their relation. 
\end{itemize}

There is a SQUARE 25010 subcategory of the Satisfaction characteristic of Quality in Use called "comfort": degree to which the user is satisfied with physical comfort.

\subsection{Fun} \label{s:Fun}
One of the main qualities of software for children is that it should be fun.
We define "fun" as the degree to which children enjoy interacting with a software product. Fun can be divided into two main subcharacteristics:

\begin{itemize}

\item Digital \textbf{Entertainment} is mainly associated with teenagers playing video games online. The interactivity of the medium allows a player to choose settings or the unfolding of
a narrative, to participate in the narrative, pursue goals, accept challenges, and experience. The study of the relation between software and children has been dominated by computer games research \cite{mayer2019computer}. 

\item \textbf{Gamification} is defined “As a way to use game elements to learn" \cite{hamari2014does}. Gamification uses game-like features including points and various levels in a way that is not meant to be mere entertainment, but to provide solutions to problems and/or to provide training, practice, and interactions that are engaging while utilizing real-world objects” \cite{de2013factors}. Gamification has been defined as a process of
enhancing services with (motivational) affordances in
order to invoke gameful experiences and further
behavioral outcomes. 
The role of gamification in general software is to add a layer that provides the same psychological experiences as games do.

Since the invention of the digital computer, games have been developed for education in various subjects, like mathematics and foreign languages, by adding a layer of gamification to subject learning, and according to \cite{McLean1978}, educational games were already popular in elementary and secondary schools in the 70's.  
Key influences on the successful use of games to support struggling readers (repetition, feedback, motivation, self‐efficacy, parental beliefs) are reported in \cite{holmes2011using}

\end{itemize}

Overall, Fun can be seen as related to the Pleasure subcategory of the Satisfaction characteristic of Quality in Use, which also includes the pleasure to use a product to satisfy such as acquiring new knowledge and skills.

\subsection{Creativity} \label{s:Creativity}
Digital creativity for children is characterized by creativity support tools and activity designs to assist users engaged in creative work. Examples of creativity measures can be found in \cite{cherry2014quantifying}. 

Digital creativity is defined as the creativity manifested in all forms that are driven by digital technologies. Digital creativity can be divided into two subcharacteristics. 

\begin{itemize} 

\item Creativity for \textbf{Storytelling}. Digital storytelling tools enable children to develop multimedia stories. As observed by \cite{rubegni2014fiabot}, digital storytelling creativity cannot be achieved only by digital device to support the creative process. The software has to be introduced into already existing practices, including the interaction between the child, the teachers, and educational processes.

\item Creativity for \textbf{Programming}. Since the public launch in May 2007, the Scratch Web site functions as a platform and  online community for digital creativity for children, with people sharing, discussing, and remixing one another's coding projects \cite{resnick2009scratch}. Paper \cite{papavlasopoulou2019exploring} explores digital creativity  for children  and proposes activities that combine art and programming for children. 

\end{itemize}

\section{Guidelines}
\label{guidelines}

We now provide two sets of preliminary guidelines, one for developers (in Section \ref{developmentGuidelines}) and the other for researchers (in Section \ref{ResearchGuidelines}).

Common to both research and development is attention to Ethical issues.
When developing for children and with children and when researching children as subjects, parents or guardians must grant practitioners and researchers consent to collect and store data. Procedures must be established in accordance with the national authorities for data. When health data are collected, one needs to be even more careful and requests for extra permissions must be addressed to health authorities. In general, data have to be anonymized and there must exist a precise plan for when to delete the data after the analysis.
Special attention has to be given to ethical issues when children are subjects of empirical investigations for software development. More refined guidelines must be defined specifically addressing the involvement of children, similarly to existing guidelines for using  university students as subjects in SE research while balancing research and educational goals \cite{carver2010checklist}. 

The EU General Data Protection Regulation (GDPR) brought new rights for European residents to have control over their online personal data. In addition, online data controllers and processors must also take new steps for ensuring personal data is secured. GDPR \footnote{https://gdpr-info.eu/}  devotes one of its 178 recitals (Recital 38 Special protection of children’s personal data). In the United States, Children's Online Privacy Protection Rule (``COPPA'') \footnote{https://www.ftc.gov/enforcement/rules/rulemaking-regulatory-reform-proceedings/childrens-online-privacy-protection-rule} imposes certain requirements on operators of websites or online services directed to children under 13 years of age, and on operators of other websites or online services that have actual knowledge that they are collecting personal information online from a child under 13 years of age.

\subsection{Guidelines for Development}
\label{developmentGuidelines}

Based on previous studies carried out about single qualities, like well-being \cite{michalsen2020family} \cite{marikken-toddlers}, we propose a preliminary set of guidelines that we outline next.

\begin{enumerate}

\item The \textbf{child and the caregivers} should be included as much as possible in the software development process, including testing. It is not uncommon nowadays that children are invited to universities and to industries to participate in coding workshops. The ideas developed by the children should be incorporated as much as possible into the software developed by the companies, especially when the companies develop software for children.

\item \textbf{Each of the qualities} (and subqualities) should be considered \textbf{in each phase of the software development process}. If a quality is regarded as not to be of primary importance for a specific software development project, the software development team should discuss and document why it is not important. Consider for example fun. It cannot be intuitive to consider fun when developing a system for safety, but studies, see for example \cite{read2002endurability}, reveal the importance of understanding and measuring fun in software systems that are devoted to children, which should be fun to use. 
\end{enumerate}





\subsection{Guidelines for Research}
\label{ResearchGuidelines}

Software engineering is a multi-disciplinary field, crossing many social and technological boundaries. Software engineering processes are studied by interdisciplinary efforts that combine technical, business, and social perspectives \cite{fitzgerald2017}.


Thus, research should be carried out in the context of a general research question that should guide future research in Software Engineering with children: 

\begin{itemize}
\item How can Software Engineering knowledge be extended to incorporate knowledge about children as stakeholders? 
\end{itemize}

This general question can be refined in several ways. For instance, if the focus in on software development, a relevant research question may be
\begin{itemize}
\item How to design processes for involving children in Software Engineering development? 
\end{itemize}

As an example, when it comes to software quality, this general question can be refined as:
\begin{itemize}
\item What are the relevant qualities of software aimed at children? 
\end{itemize}

We now introduce a few guidelines with the long term \textbf{aim} to develop validated interdisciplinary knowledge about \textbf{software quality and children} to help answer these research questions (and other related ones). 

\begin{enumerate}

\item The building of software, whether especially conceived for children or not, must be studied from the point of view of various stakeholders, such as:
\begin{itemize}
\item children of various ages, skills, and different social and cultural contexts and their caregivers. 
\item software engineers who work in software projects that develop software for children. 
\item software engineers who work in software projects that develop software for all, since, as observed before, children use both specific software and software made for general users. 
\end{itemize}

\item Researchers must be aware of the fact that technical aspects, although necessary, represent only a part of the set of problems that need to be addressed. To understand processes that develop and evolve software systems with children as stakeholders, researchers need to investigate tools and also the social and cognitive processes surrounding them. Research must draw from several different sources and disciplines.

\item Research in this field cannot be purely theoretical or speculative, but it must be carried out via empirical studies. It will be necessary to carry out systematic collections and analyses of empirical data to develop validated knowledge about why and how organizations, teams, and individual software engineers
develop software \cite{easterbrook2008selecting} when children are, or should be considered, relevant stakeholders. 

\item Data collection should be carried out for specific goals and in the framework of a quality model like the one we proposed in Section \ref{s:Qualities}. For each quality, carefully designed templates should be used to gather information from each stakeholder about: 
\begin{itemize}
\item characteristics of the software under development; 
\item characteristics of the software process in use, like agile, extreme programming, etc.; 

\item the intention of children to participate in Software Engineering activities;  
\item  the intention of software engineers to integrate children in the Software Engineering activities; 

\item relations between qualities and software development phases (Like for example, "in which phase do you work with mental well-being issues?") 
\item the reciprocal relations between the network of qualities and their sub qualities
\item the relative importance of the qualities, like fun can be perceived as more important by small children, than by adolescents, or software developers.  
\end{itemize}

It will be important to translate these questions into a language that is understandable for children, see for example  \cite{read2002endurability} for tools to elicit information from young children. Study  \cite{marikken-toddlers} reports about  data collection about the interaction of toddlers and their care givers with researchers and medical personnel. They have used Affinity diagram to structure the elicited knowledge.

\item More generally, it will be important to define what type of Software Engineering knowledge and education children need to be able to effectively participate in Software Engineering processes.

\end{enumerate}

\section{Conclusions and Future Work}
\label{conclusions}


We have proposed a model that puts children goals and well-being as an integral part of the software engineering processes, so that the children who use software systems will be offered new possibilities to influence the future of software systems and they will be made aware of threads that can be caused by software systems.

 There is no common definition about how to characterize an individual as a child, given her age. 
 Age-related development periods and examples of defined intervals include (according to \cite{10.5555/2017483.2017487}): newborn (ages 0–4 weeks); infant (ages 4 weeks – 1 year); toddler (ages 12 months-24 months); preschooler (ages 2–5 years); school-aged child (ages 6–12 years); adolescent (ages 13–19). In this work, we have studied children from the perspective of their relation to technology and we have presented related work and background that spans from research about toddlers and technology, like in \cite{marques2014systematic} to research with adolescents, like in \cite{michalsen2020family}.
 A limitation of our work is that we have not gone in depth into the different age categories and this distinction by age has to be addressed by further work.

 We have reviewed studies devoted to understand single qualities, like creativity and guidelines to develop for one quality, but the qualities and the guidelines have not been evaluated in its wholeness yet.
 The proposed characteristics and sub-characteristics have to be validated by setting up systematic investigations of the literature and of the practice. We have proposed a road map to set up empirical investigations to grasp the perspective of the different stakeholders, including software engineers, children, care givers. The proposed road map (four main qualities and guidelines for practice and research) will enable researchers to set up the empirical investigations interventions in SE with children.


There is consensus in the SE literature about the distinction between qualities and the respective activities to achieve the given quality, like  "maintenance" is an activity, but "maintainability" is the corresponding quality. On the contrary, in existing literature about software for children, Gamification is used for both the quality of the software and the activity of gamifying the software. The same applies to creativity. Further work will refine our model and propose increased understanding and better definitions of the qualities and the respective activities. We will also explore the relationships between existing quality models and the qualities of interest for children. For instance, while the Confidentiality of Security in the SQUARE 25010 standard can be somewhat mapped into our preliminary quality model, the role and relevance (if any) of the other subcharacteristics, i.e, integrity, non-repudiation, accountability, and authenticity, still need to be investigated. In general, further work will establish Software Engineering with children as a sub discipline of Software Engineering with a specific terminology, models, techniques, and methods.

\bibliographystyle{unsrt}

\bibliography{bib}

\begin{thebibliography}{10}

\bibitem{holmes2011using}
Wayne Holmes.
\newblock {Using game-based learning to support struggling readers at home}.
\newblock {\em Learning, Media and Technology}, 36(1):5--19, 2011.

\bibitem{hagen2016gameplay}
Kristoffer Hagen, Konstantinos Chorianopoulos, Alf~Inge Wang, Letizia Jaccheri,
  and Stian Weie.
\newblock {Gameplay as exercise}.
\newblock In {\em {Proceedings of the 2016 CHI Conference Extended Abstracts on
  Human Factors in Computing Systems}}, pages 1872--1878, 2016.

\bibitem{papavlasopoulou2017empirical}
Sofia Papavlasopoulou, Michail~N Giannakos, and Letizia Jaccheri.
\newblock {Empirical studies on the Maker Movement, a promising approach to
  learning: A literature review}.
\newblock {\em Entertainment Computing}, 18:57--78, 2017.

\bibitem{resnick2009scratch}
Mitchel Resnick, John Maloney, Andr{\'e}s Monroy-Hern{\'a}ndez, Natalie Rusk,
  Evelyn Eastmond, Karen Brennan, Amon Millner, Eric Rosenbaum, Jay Silver,
  Brian Silverman, et~al.
\newblock {Scratch: programming for all}.
\newblock {\em Communications of the ACM}, 52(11):60--67, 2009.

\bibitem{unicef2019}
Office of~Global~Insight and Policy.
\newblock {Workshop report: AI and child rights policy.}
\newblock Technical report, {United Nations Children Fund}, 2019.

\bibitem{hamdan2013protecting}
Zainab Hamdan, Iman Obaid, Asma Ali, Hanan Hussain, Amala~V Rajan, and Jinesh
  Ahamed.
\newblock {Protecting teenagers from potential internet security threats}.
\newblock In {\em {2013 International Conference on Current Trends in
  Information Technology (CTIT)}}, pages 143--152. IEEE, 2013.

\bibitem{worldreport}
{Margie Peden et al.}
\newblock {World report on child injury prevention.}
\newblock Technical report, {World Health Organization}, 2008.

\bibitem{vandewater2004linking}
Elizabeth~A Vandewater, Mi-suk Shim, and Allison~G Caplovitz.
\newblock {Linking obesity and activity level with children's television and
  video game use}.
\newblock {\em Journal of adolescence}, 27(1):71--85, 2004.

\bibitem{hourcade2008interaction}
Juan~Pablo Hourcade et~al.
\newblock {Interaction design and children}.
\newblock {\em Foundations and Trends{\textregistered} in Human--Computer
  Interaction}, 1(4):277--392, 2008.

\bibitem{ISO25010:2011}
{ISO/IEC 25010}.
\newblock {\em {ISO}/{IEC} 25010:2011, Systems and software engineering —
  Systems and software Quality Requirements and Evaluation (SQuaRE) — System
  and software quality models}.
\newblock ISO/IEC, 2011.

\bibitem{FentonBiemanBOOK2014}
Norman~E. Fenton and James~M. Bieman.
\newblock {\em {Software Metrics: A Rigorous and Practical Approach, Third
  Edition}}.
\newblock Chapman \& Hall/CRC Innovations in Software Engineering and Software
  Development Series. Taylor \& Francis, 2014.

\bibitem{MorascaESEM2009}
Sandro Morasca.
\newblock {A probability-based approach for measuring external attributes of
  software artifacts}.
\newblock In {\em Proceedings of the 2009 3rd International Symposium on
  Empirical Software Engineering and Measurement}, ESEM '09, Lake Buena Vista,
  FL, USA, October 15-16, 2009, pages 44--55, Washington, DC, USA, 2009. IEEE
  Computer Society.

\bibitem{Giannakas2016}
Filippos Giannakas, Georgios Kambourakis, Andreas Papasalouros, and Stefanos
  Gritzalis.
\newblock {Security Education and Awareness for K-6 Going Mobile}.
\newblock {\em International Journal of Interactive Mobile Technologies
  (iJIM)}, 10(2):41--48, 2016.

\bibitem{lwin2008protecting}
May~O Lwin, Andrea~JS Stanaland, and Anthony~D Miyazaki.
\newblock Protecting children's privacy online: How parental mediation
  strategies affect website safeguard effectiveness.
\newblock {\em Journal of Retailing}, 84(2):205--217, 2008.

\bibitem{marikken-toddlers}
Marikken H\o{}iseth, Michail~N. Giannakos, Ole~A. Alsos, Letizia Jaccheri, and
  Jonas Asheim.
\newblock {Designing Healthcare Games and Applications for Toddlers}.
\newblock In {\em Proceedings of the 12th International Conference on
  Interaction Design and Children}, IDC ’13, page 137–146, New York, NY,
  USA, 2013. Association for Computing Machinery.

\bibitem{michalsen2020family}
H~Michalsen, SC~Wangberg, A~Anke, G~Hartvigsen, L~Jaccheri, and C~Arntzen.
\newblock Family members and health care workers' perspectives on motivational
  factors of participation in physical activity for people with intellectual
  disability: A qualitative study.
\newblock {\em Journal of Intellectual Disability Research}, 64(4):259--270,
  2020.

\bibitem{gomez2019digital}
Javier Gomez~Escribano, Maria~Letizia Jaccheri, Manolis Maragoudakis, and
  Kshitij Sharma.
\newblock {Digital storytelling for good with Tappetina game}.
\newblock {\em Entertainment Computing Journal}, 2019.

\bibitem{mayer2019computer}
Richard~E Mayer.
\newblock {Computer games in education}.
\newblock {\em Annual review of psychology}, 70:531--549, 2019.

\bibitem{hamari2014does}
Juho Hamari, Jonna Koivisto, and Harri Sarsa.
\newblock {Does gamification work?--a literature review of empirical studies on
  gamification}.
\newblock In {\em {2014 47th Hawaii international conference on system
  sciences}}, pages 3025--3034. IEEE, 2014.

\bibitem{de2013factors}
Penny De~Byl.
\newblock {Factors at play in tertiary curriculum gamification}.
\newblock {\em International Journal of Game-Based Learning (IJGBL)},
  3(2):1--21, 2013.

\bibitem{McLean1978}
H.~W. McLean.
\newblock {Are simulations and games really legitimate?}
\newblock {\em Audiovisual Instruction}, 23(7):12--13, 1978.

\bibitem{cherry2014quantifying}
Erin Cherry and Celine Latulipe.
\newblock {Quantifying the creativity support of digital tools through the
  creativity support index}.
\newblock {\em ACM Transactions on Computer-Human Interaction (TOCHI)},
  21(4):1--25, 2014.

\bibitem{rubegni2014fiabot}
Elisa Rubegni and Monica Landoni.
\newblock Fiabot! design and evaluation of a mobile storytelling application
  for schools.
\newblock In {\em Proceedings of the 2014 conference on Interaction design and
  children}, pages 165--174, 2014.

\bibitem{papavlasopoulou2019exploring}
Sofia Papavlasopoulou, Michail~N Giannakos, and Letizia Jaccheri.
\newblock {Exploring children's learning experience in constructionism-based
  coding activities through design-based research}.
\newblock {\em Computers in Human Behavior}, 2019.

\bibitem{carver2010checklist}
Jeffrey~C Carver, Letizia Jaccheri, Sandro Morasca, and Forrest Shull.
\newblock {A checklist for integrating student empirical studies with research
  and teaching goals}.
\newblock {\em Empirical Software Engineering}, 15(1):35--59, 2010.

\bibitem{read2002endurability}
Janet~C Read, Stuart MacFarlane, and Chris Casey.
\newblock {Endurability, engagement and expectations: Measuring children’s
  fun}.
\newblock In {\em {Interaction design and children}}, volume~2, pages 1--23.
  Shaker Publishing Eindhoven, 2002.

\bibitem{fitzgerald2017}
Brian Fitzgerald and Klaas-Jan Stol.
\newblock {Continuous software engineering: A roadmap and agenda}.
\newblock {\em Journal of Systems and Software}, 123:176--189, 2017.

\bibitem{easterbrook2008selecting}
Steve Easterbrook, Janice Singer, Margaret-Anne Storey, and Daniela Damian.
\newblock Selecting empirical methods for software engineering research.
\newblock In {\em Guide to advanced empirical software engineering}, pages
  285--311. Springer, 2008.

\bibitem{10.5555/2017483.2017487}
Maung~K. Sein, Ola Henfridsson, Sandeep Purao, Matti Rossi, and Rikard
  Lindgren.
\newblock {Action Design Research}.
\newblock {\em MIS Q.}, 35(1):37–56, March 2011.

\bibitem{marques2014systematic}
Ma{\'\i}ra~R Marques, Alcides Quispe, and Sergio~F Ochoa.
\newblock {A systematic mapping study on practical approaches to teaching
  software engineering}.
\newblock In {\em {2014 IEEE Frontiers in Education Conference (FIE)
  Proceedings}}, pages 1--8. IEEE, 2014.

\end{thebibliography}

\end{document}